\documentstyle[12pt]{article}
\begin{document}
\tolerance=5000
\def\eg{{\it e.g.}\ }
\def\ie{{\it i.e.}\ }
\def\det{{\rm det}}
\def\tr{{\rm tr}}
\def\Tr{{\rm Tr}}
\def\e{{\rm e}}

\begin{flushright}
OCHA-PP-61 \\
NDA-FP-20 \\
%hep-th/950xxxx
June (1995)
\end{flushright}
\vfill
\begin{center}
{\large\bf Two-Form Gravity and the Generation
of Space-Time}

\vfill

{\sc Miyuki KATSUKI,\footnote{Present Address :
Fujitsu Ltd., Kawasaki, Kanagawa, 211 JAPAN}
Shin'ichi NOJIRI${^\spadesuit}$
and Akio SUGAMOTO}

\vfill

{\it Department of Physics,  Ochanomizu University}

{\it 1-1, Otsuka 2, Tokyo 112, JAPAN}

\vfill

{\it ${}^\spadesuit$Department of Mathematics and
Physics}

{\it National Defence Academy}

{\it Hashirimizu, Yokosuka 139, JAPAN}

\vfill

{\bf ABSTRACT}
\end{center}

In the framework of the two-form gravity, which
is classically equivalent to the Einstein gravity,
the one-loop effective potential
for the conformal factor of metric is calculated in
the finite
volume and in the finite temperature by choosing a
temporal gauge condition.
There appears a quartically divergent term which cannot
be removed by the renormalization of the cosmological
term and we find there is only one non-trivial minimum
in the effective potential.
If the cut-off scale has a physical meaning, \eg the
Planck scale coming from string theory, this minimum
might explain why the space-time is generated, \ie why
the classical metric has a non-trivial value.

\newpage

\section{Introduction}

Although the classical theory of the Einstein gravity
is simple and successfully describes the nature,
we encounter the many serious problems when
we try to construct a quantum theory of gravity.
One of these problems is that the action of the Einstein
gravity is not renormalizable.
This might suggest that the Einstein gravity theory is
an effective theory obtained from a more fundamental
theory, \eg string theory.
Two-form gravity theory is known to be classically
equivalent to the Einstein gravity theory and is obtained
from a topological field theory, which is called BF
theory  \cite{sch},
by imposing constraint conditions \cite{pl}.
The characteristic feature of the BF theory is that
the system has the Kalb-Ramond symmetry \cite{kara},
which is a large
local symmetry. The Kalb-Ramond symmetry can be considered
to reflect the stringy structure of the fundamental gravity
theory \cite{kkns}.

In this paper, we calculate, in the framework of the
two-form gravity, the one-loop effective potential
for the conformal factor of the metric in the finite
volume and in the finite temperature by choosing a
temporal gauge fixing condition.
In the effective potential, there appears a quartically
divergent term which cannot be removed by the
renormalization of the cosmological term.
We also find that there is only one non-trivial minimum
in the effective potential, which might explain why
the space-time is generated, \ie why the classical metric
has a non-trivial value if the cut-off scale has a physical
meaning, \eg the Planck scale coming from string theory.

In the next section, we explain the action of the two-form
gravity and clarify the gauge symmetries of the system.
In section 3, we fix the gauge symmetries and expand the
action around a classical solution. The measures which keeps
the gauge symmetries and the action of the ghost fields are
given in section 4. In section 5, the one-loop effective
potential for the conformal factor of metric is calculated.
The last section is devoted to the summary.

\section{Two-Form Gravity Action}

We start with the following action which describes a
topological field theory called BF theory \cite{sch}:
\begin{equation}
\label{bf}
     S = \int d^4x \frac{1}{2} \epsilon^{\mu\nu\lambda\rho}
\left( \: B^a_{\mu\nu}(x)R^a_{\lambda\rho}(x) +
\bar B^a_{\mu\nu}(x)  \bar R^a_{\lambda\rho}(x)  \:
\right)\ .
\end{equation}
Here $B^a_{\mu\nu}(x)$ and $\bar B^a_{\mu\nu}(x)$ are
two-form fields and $R^a_{\mu\nu}(x)$ and
$\bar R^a_{\mu\nu}(x)$ are $SU(2)$ field strength ($a$
denotes an $SU(2)$ index $a=1,2,3$) given by
the $SU(2)$ gauge fields $A$ and $\bar A$:
\begin{equation}
\label{curvature}
R^a_{\mu\nu}=\partial_\mu A^a_\nu - \partial_\nu A_\mu^a
+g\epsilon^{abc}A^b_\mu A^c_\nu\ , \hskip 0.5cm
\bar R^a_{\mu\nu}=\partial_\mu\bar A_\nu
- \partial_\nu\bar A_\mu
+g\epsilon^{abc}\bar A^b_\mu\bar A^c_\nu \ .
\end{equation}
Here $g$ is a gauge coupling constant.
The gauge fields $A$ and $\bar A$ are identified
with spin connections.
The action (\ref{bf}) has $SU(2)\times SU(2)$
gauge symmetry corresponding to local Lorentz symmetry
$SO(4)\sim SU(2)\times SU(2)$. Besides $SU(2)\times SU(2)$
gauge transformation, the action is invariant under
the Kalb-Ramond symmetry transformation \cite{kara}:
\begin{equation}
\label{kr1}
\left\{
\matrix{
A^a_\mu & \rightarrow & A^a_\mu \cr
B^a_{\mu\nu} & \rightarrow & B^a_{\mu\nu} +
\nabla^{ab}_\mu \Lambda^b_\nu - \nabla^{ab}_\nu \Lambda^b_\mu
\ , \cr}\right.
\end{equation}
\begin{equation}
\label{kr2}
\left\{
\matrix{
A^a_\mu & \rightarrow & A^a_\mu \cr
B^a_{\mu\nu} & \rightarrow & B^a_{\mu\nu} +
\nabla^{ab}_\mu \Lambda^b_\nu - \nabla^{ab}_\nu \Lambda^b_\mu
\ .\cr } \right.
\end{equation}
Here the covariant derivative $\nabla^{ab}_\mu$ is defined by
$ \: \nabla^{ab}_\mu\Lambda^b_\nu = \partial_\mu \Lambda^a_\nu +
\epsilon^{abc}A^b_\mu\Lambda^c_\nu $.

The action (\ref{bf}) is known to be equivalent to the Einstein
action if we impose the following constraints \cite{pl}:
\begin{eqnarray}
\label{cons1}
 \epsilon^{\mu\nu\lambda\rho} (B^a_{\mu\nu}B^b_{\lambda\rho} -
\frac{1}{3}\delta^{ab}B^c_{\mu\nu}B^c_{\lambda\rho}) &=& 0 \ ,
\\
\label{cons2}
\epsilon^{\mu\nu\lambda\rho} (\bar B^a_{\mu\nu}
\bar B^b_{\lambda\rho} -
\frac{1}{3}\delta^{ab}\bar B^c_{\mu\nu}\bar B^c_{\lambda\rho})
&=& 0 \ , \\
\label{cons3}
\epsilon^{\mu\nu\lambda\rho} B^a_{\mu\nu}
\bar B^b_{\lambda\rho} &=& 0
\ .
\end{eqnarray}
In the following, we only consider the chiral part of the theory,
which is given by $B^a_{\mu\nu}$ and $A^a_\mu$, since
the anti-chiral part given by $\bar B^a_{\mu\nu}$
and $\bar A^a_\mu$ is a copy of chiral part and
the two parts are decoupled with each other.

We impose the constraint (\ref{cons1}) by a multiplier field
$\phi^{ab}$:
\begin{equation}
  S = \int d^4x \frac{1}{2} \epsilon^{\mu\nu\lambda\rho}
\left( B^a_{\mu\nu}(x)R^a_{\lambda\rho}(x) +
\phi^{ab}(x)(B^a_{\mu\nu}B^b_{\lambda\rho} -
\frac{1}{3}\delta^{ab}B^c_{\mu\nu}B^c_{\lambda\rho})\right)
\label{chiral1}\ .
\end{equation}
The action (\ref{chiral1}) is not invariant under the Kalb-Ramond
transformation (\ref{kr1}) but we can keep the symmetry
by introducing a new field $G^a_\mu$, which is transformed by
\begin{equation}
\label{kr3}
 \delta G^a_\mu = \Lambda^a_\mu
\end{equation}
and by modifying the action:
\begin{eqnarray}
\label{chiral2}
    S &=& \int d^4x \epsilon^{\mu\nu\rho\sigma}
\Bigl[  B^a_{\mu\nu}
R^a_{\rho\sigma}  \nonumber \\
& & + \phi^{ab}\{(B^a_{\mu\nu} - \nabla_\mu G^a_\nu+
\nabla_\nu G^a_\mu)(B^b_{\rho\sigma} -
\nabla_\rho G^b_\sigma + \nabla_\sigma G^b_\rho) \\
& & \ \ - \frac{1}{3} \delta^{ab}
(B^c_{\mu\nu} - \nabla_\mu G^c_\nu+\nabla_\nu G^c_\mu)
(B^c_{\rho\sigma} - \nabla_\rho G^c_\sigma + \nabla_\sigma
G^c_\rho)  \} \Bigr] \ . \nonumber
\end{eqnarray}
We can also consider the cosmological term
\begin{equation}
\label{cosmo}
S_{\rm cosmo} =\int d^4x \,
\Lambda\epsilon^{\mu\nu\rho\sigma}
(B^a_{\mu\nu} - \nabla_\mu G^a_\nu+
\nabla_\nu G^a_\mu)(B^a_{\rho\sigma} -
\nabla_\rho G^a_\sigma + \nabla_\sigma G^a_\rho) \ .
\end{equation}

\section{Gauge Fixing}

The actions (\ref{chiral2}) and (\ref{cosmo}) have
the following local symmetries: 1. the Kalb-Ramond
symmetry, 2. $SU(2)$ gauge symmetry, which is the
chiral part of the local Lorentz symmetry, 3. general
covariance. The condition
\begin{equation}
\label{gauge1}
G^a_\mu=0
\end{equation}
fixes the Kalb-Ramond symmetry.
This gauge condition (\ref{gauge1}) does not generate
any ghost action. We also fix the $SU(2)$ gauge symmetry
by choosing the temporal gauge condition:
\begin{equation}
\label{gauge2}
A^a_4=0\ .
\end{equation}
In the usual gauge theory, the ghost corresponding
to the temporal gauge (\ref{gauge2}) does not
contribute to the physical amplitude since the
corresponding Jacobian is a $c$-number. In the
two-form gravity theory, however, the Jacobian
depends on the space-time metric, as we will
see later, and there are
contributions from the ghost fields
to the physical amplitude.

When the cosmological constant $\Lambda$ vanishes
$\Lambda=0$, a solution of the equations of motion
derived from the action (\ref{chiral2}) is given by
\begin{equation}
\label{sol}
B^a_{\mu\nu} = \varphi^2\eta^a_{\mu\nu}\ ,
\hskip 1cm {\rm other\ fields}=0
\end{equation}
Here $\varphi$ is a constant, which cannot be
determined from the equations of motion, and
$\eta^a_{\mu\nu}$
is the 't Hooft symbol \cite{thft2} which is defined by,
\begin{equation}
\label{thooft}
\eta^a_{\mu\nu} = \left\{
\matrix{
\epsilon^{a\mu\nu} \: & &\mu,\nu = 1,2,3 \cr
\delta^a_\mu \:& &\: \nu=4\:\: \cr
-\delta^a_\nu \:& &\: \mu=4\:\: \cr
} \right.
\end{equation}
In the following, we calculate the radiative
correction \cite{lnu} of the effective potential for
$\varphi$ by expanding the action around the
solution (\ref{sol}) and keeping the quadratic terms:
\begin{eqnarray}
\label{action1}
S &\sim & \int d^4x [ 2g\varphi^2(A^a_aA^b_b
- A^a_bA^b_a) -
2\epsilon^{ijk} b^a_{ij}\partial_4 A^a_k +
4\epsilon^{ijk}b^a_{i4}\partial_j A^a_k \nonumber \\
      & & \: + \varphi^2 \phi^{ab}
\{2b^a_{ij}\epsilon^{ijb} +
2b^b_{ij}\epsilon^{ija} + 4 b^a_{b4} + 4b^b_{a4} \nonumber \\
& & \: - \frac{1}{3} \delta^{ab} ( 4b^c_{ij}\epsilon^{ijc}
+  b^c_{c4}) \} ]\ .
\end{eqnarray}
Here $b^a_{\mu\nu}$ is defined by
$ B^a_{\mu\nu} = \varphi^2\eta^a_{\mu\nu}
+ b^a_{\mu\nu} $.
The action (\ref{action1}) is invariant under
a global $SU(2)$ symmetry. The global $SU(2)$ symmetry
is the diagonal part of the $SU(2)$ symmetry coming
from the $SU(2)$ gauge symmetry and the $SU(2)\sim SO(3)$
symmetry of the rotation in the three dimensional space.
We can now decompose $b^a_{\mu\nu}$ and $A^a_b$
into the diagonal $SU(2)$ spin 0, 1 and 2
components:
\begin{equation}
\label{deco}
\left\{
\matrix{
  2b^a_{ij}\epsilon^{ijb} &=& \delta^{ab} h + \epsilon^{abc} h^c
+ h^{ab} \cr
  4b^a_{i4} &=& \delta^{ai}e + \epsilon^{aic} e^c + e^{ai} \cr
   A^a_b &=& \delta^{ab}a + \epsilon^{abc}a^c +a^{ab} \cr
} \right. \ .
\end{equation}
By partial integration and the redefinition of the
fields, the action (\ref{action1}) is rewritten by,
\begin{eqnarray}
\label{action2}
S &\sim& \int d^4 x [ 12g\varphi^2
\tilde{a}^2 +4g\varphi^2\tilde{a}^a
\tilde{a}^a
- 2g \varphi^2\tilde{a}^{ab}\tilde{a}^{ab}
\nonumber \\
  & & -\frac{3}{16g\varphi^2}
(\partial_4\tilde{h})^2
 - \frac{1}{4g\varphi^2}(\partial_4\tilde{h}^a)^2
+ \frac{1}{8g\varphi^2}(\partial_4\tilde{h}^{ab})^2
\nonumber \\
  & & - 8g\varphi^6(\partial_4^{-1}
\tilde{\phi}^{ab})^2 \nonumber \\
  & & +\frac{1}{8g\varphi^2}
\{(\: \frac{1}{2} \partial_ae^b +
\partial_be^a - \frac{2}{3} \delta^{ab}\partial_ce^c)
\nonumber \\
& & + \frac{1}{2} (\epsilon^{dcb} \partial_c e^{ad}
+ \epsilon^{dca}\partial_c e^{bd} )
- \partial_4 e^{ab}\: \}^2  \ .
\end{eqnarray}
The redefined fields are given by
\begin{eqnarray}
\label{redef}
 \tilde{a} &=& a
+  \frac{1}{8g\varphi^2}\partial_4 h -
\frac{1}{12g\varphi^2}\partial_a e^a  \nonumber \\
 \tilde{a}^a &=& a^a
+ \frac{1}{4g\varphi^2} \partial_4a^a +
\frac{1}{4g\varphi^2}\partial_ae
+ \frac{1}{8g\varphi^2}
\epsilon^{abc} \partial_b e^c
- \frac{1}{16g\varphi^2}
(\partial_c e^{ca} + \partial_c e^{ac}) \nonumber \\
 \tilde{a}^{ab} &=& a^{ab}
-\frac{1}{4g\varphi^2} \partial_4 h^{ab}
- \frac{1}{4g\varphi^2} \Bigl( \partial_a e^b
+ \partial_b e^a -
\frac{2}{3} \delta^{ab}\partial_c e^c \Bigr)
\nonumber \\
&& + \frac{1}{4g\varphi^2} (\epsilon^{dcb}
\partial_c e^{ad}  + \epsilon^{dca}\partial_c e^{bd} )
\nonumber \\
%%%%%%%%%%%%%%%%%%
 \tilde{h} &=& h - \frac{2}{3} \partial_4^{-1}
\partial_a e^a \nonumber \\
 \tilde{h} ^a &=& h^a + \partial_4^{-1}
\Bigl\{ \partial_a e +
\frac{1}{2} \epsilon^{abc}\partial_b e^c
- \frac{1}{4} (
\partial_c e^{ca}+ \partial_c e^{ac} ) \Bigr\}
\nonumber \\
 \tilde{h}^{ab} &=& h^{ab} -\partial_4^{-1}
\Bigl\{ \frac{1}{2} \Bigl[
\partial_a e^b + \partial_b e^a
- \frac{2}{3} \delta^{ab}
\partial_c e^c -( \epsilon^{dcb}
\partial_c e^{ad} + \epsilon^{dca}
\partial_ce^{bd}) \Bigr] \nonumber \\
&& -8g \varphi^4
\partial_4^{-1} \phi^{ab} \Bigr\} \nonumber \\
%%%%%%%%%%%%%%%%%%%%%%%%%%%%%%%%%%%%%%%%%%%%%%%%%%%%%%%%%%
 \tilde{\phi}^{ab} &=& \phi^{ab} - \partial_4 \Bigl\{
\frac{1}{16g\varphi^4} \Bigl( \partial_a e^b
+ \partial_b e^a -
\frac{2}{3} \delta^{ab} \partial_c e^c
- \epsilon^{dcb}
\partial_c e^{ad} - \epsilon^{dca}
\partial_c e^{bd} \Bigr) \nonumber \\
&& + \frac{1}{8g\varphi^4}\partial_4 e^{ab} \Bigr\} \ .
\end{eqnarray}

Under the general coordinate transformation
$\delta x^\mu =\epsilon^\mu$, $ b^a_{\mu\nu}$
transform as
\begin{equation}
\label{gct}
\delta b^a_{\mu\nu} = \varphi^2
( \partial_\mu \epsilon^\rho
\eta^a_{\rho\nu} + \partial_\nu
\epsilon^\rho \eta^a_{\mu\rho} )
 + \epsilon_\rho \partial_\rho b^a_{\mu\nu}
+ \partial_\mu
\epsilon^\rho b^a_{\rho\nu}+ \partial_\nu
\epsilon^\rho
b^a_{\mu\rho}
\end{equation}
since $\delta B^a_{\mu\nu} = \epsilon^\rho
\partial_\rho B^a_{\mu\nu} + \partial_\mu
\epsilon^\rho
B^a_{\rho\nu} + \partial_\nu \epsilon^\rho
B^a_{\mu\rho} $.
Note that there appear inhomogeneous terms in
Eq.(\ref{gct}). Therefore we can fix the local
symmetry of the general coordinate transformation
by choosing the following conditions:
\begin{equation}
\label{gauge3}
e=e^a=0\ .
\end{equation}
This gauge condition is a kind of temporal gauge since
$e$ and $e^a$ are parts of $b^a_{i4}$ (\ref{deco}).
Then the action has the following form (the actions
of the ghost fields are given in the next section.):
\begin{eqnarray}
\label{action3}
  S &\sim& \int d^4 x [ 12g\varphi^2 \tilde{a}^2
+4g\varphi^2\tilde{a}^a
\tilde{a}^a - 2g \varphi^2\tilde{a}^{ab}
\tilde{a}^{ab} \nonumber \\
  & & -\frac{3}{16g\varphi^2}
(\partial_4\tilde{h})^2
 - \frac{1}{4g\varphi^2}(\partial_4\tilde{h}^a)^2
+ \frac{1}{8g\varphi^2}(\partial_4\tilde{h}^{ab})^2
\nonumber \\
  & & - 8g\varphi^6(\partial_4^{-1}
\tilde{\phi}^{ab})^2 \nonumber \\
  & & + \frac{1}{32g\varphi^2} \{(\epsilon^{dcb}
\partial_c e^{ad}
+ \epsilon^{dca}\partial_c e^{bd} )
-2 \partial_4 e^{ab} \}^2 ]\ .
\end{eqnarray}
After the partial integration, the last term can
be decomposed orthogonally as follows:
\begin {eqnarray}
\label{deco2}
& & \int d^4 x [ \{(\epsilon^{dcb}
\partial_c e^{ad}
+ \epsilon^{dca}\partial_c e^{bd} )
-2 \partial_4 e^{ab} \}^2 ] \nonumber \\
&=& \int d^4 x [ e^{ab}  \{-(2 \triangle +
4 \partial_4 ^2) I  +  A
+ B \} e^{cd}] \nonumber \\
    &=& \int d^4 x [ e^{ab} \{ -4\partial_4^2 M_1
- (\triangle + 4
\partial_4^2) M_2 - 4 \partial_4^2 M_3 \nonumber \\
& & - (4\triangle + 4
\partial_4^2) M_4 \}_{(ab)(cd)} e^{cd}] \\
    & & ( \triangle = \sum_{a=1}^{3}
\partial_a \partial_a \: \:
, \: \: M_i\:M_j = \delta_{ij}M_i ) \nonumber
\end{eqnarray}
$M_i$'s are given in Appendix A.

\section{Measure and Ghost Actions}

We now determine the measures of the fields,
which preserves the $SU(2)$ gauge symmetry
and the general covariance,  by
\begin{eqnarray}
\label{meas1}
( \delta B^a_{\mu\nu} )^2 &=&\int d^4x
\epsilon^{\mu\nu\rho
\sigma } \delta B^a_{\mu\nu}
\delta B^a_{\rho\sigma} \nonumber \\
   ( \delta A^a_\mu )^2 &=& \int d^4x
\epsilon^{\mu\nu\rho\sigma}
\epsilon^{abc} B^a_{\mu\nu} \delta A^b_
\rho \delta A^c_\sigma \nonumber \\
   ( \delta \phi^{ab} ) ^2 &=& \int d^4x
\epsilon^{\mu\nu\rho\sigma}
 B^a_{\mu\nu} B^a_{\rho\sigma} \delta
\phi^{bc} \delta \phi^{bc}\ .
\end{eqnarray}
In the following, we impose the periodic
boundary conditions for $x^i$ with
the periods $L$:
\begin{equation}
\label{prd}
\int_{L^3} d^3x \equiv
\int_{-\frac{L}{2}}^{+\frac{L}{2}} dx
\int_{-\frac{L}{2}}^{+\frac{L}{2}} dy
\int_{-\frac{L}{2}}^{+\frac{L}{2}} dz
\end{equation}
 Since  $ B^a_{\mu\nu} \sim
\varphi^2 \eta^a_{\mu\nu} $ {\it i.e.}
$ e^A_\mu \sim \varphi \delta^A_\mu $ ($A=1 , \cdots , 4$),
we find $ds^2 \sim \varphi^2 \sum_{\mu =1 , \cdots , 4}
dx^\mu dx^\mu$.
Therefore the volume of the universe is given by
$(\varphi L)^3$, which can be determined if
the effective potential of $\varphi$ has non-trivial
minimum.
By defining new coordinates
$\tilde x^\mu$ by $ x^\mu = \varphi^{-1}
\tilde{x}^\mu$, which gives $ds^2 \sim
\sum_{\mu =1,\cdots ,4} dx^\mu dx^\mu$, the measures
in Eq.(\ref{meas1}) can be rewritten by
\begin{eqnarray}
\label{meas2}
   ( \delta B^a_{\mu\nu} )^2 &\sim& \varphi^{-4}
\int _{(\varphi L)^3}
d^3\tilde{x} \int _0^{\frac{1}{kT}} d\tilde{x}^4
\epsilon^{\mu\nu\rho\sigma} \delta B^a_{\mu\nu}
\delta B^a_{\rho
\sigma} \nonumber \\
   ( \delta A^a_\mu )^2 &\sim & \varphi ^{-2}
\int _{(\varphi L)^3}
d^3\tilde{x} \int _0^{\frac{1}{kT}} d\tilde{x}^4
\epsilon^
{\mu\nu\rho\sigma}\epsilon^{abc} \eta^a_{\mu\nu}
\delta A^b_\rho
\delta A^c_\sigma \nonumber \\
   ( \delta \phi^{ab} )^2 &\sim &
\int _{(\varphi L)^3}d^3\tilde{x}
\int _0^{\frac{1}{kT}} d\tilde{x}^4
\epsilon^{\mu\nu\rho\sigma}\eta^a
_{\mu\nu} \eta^a_{\rho\sigma} \delta \phi^{bc}
\delta \phi^{bc}
\end{eqnarray}
Here we have replaced $ B^a_{\mu\nu}$ by $\varphi^2
\eta^a_{\mu\nu} $ and neglected higher order terms
with respect to the power of fields.
We also impose the periodic boundary
condition for $\tilde x^4$ with period $1/kT$ ($k$ is
the Boltzman constant.), \ie we are now considering
the field theory in the finite temperature $T$.

If we redefine the fields by
\begin{eqnarray}
      \hat{B}^a_{\mu\nu} &=& \varphi^{-2}
B^a_{\mu\nu}\:\:(
\rightarrow e,h) \ , \\
      \hat{A}^a_\mu &=& \varphi^{-1} A^a_\mu
\:\:(\rightarrow a) \ , \\
      \hat{\phi}^{ab} &=& \phi^{ab} \ ,
\end{eqnarray}
the $\varphi$ dependence in the measures is
absorbed and the action (\ref{action3}) has the
following form
\begin{eqnarray}
\label{action4}
   S &=& \int _{(\varphi L)^3} d^3 \tilde{x}
\int_0^{\frac{1}{kT}} d\tilde{x}^4  \nonumber \\
   & & \:\times \Bigl[ 12g\hat{a}^2 +4g\hat{a}^a
\hat{a}^a - 2g \hat{a}^{ab}\hat{a}^{ab}
\nonumber \\
   & & \: -\frac{3}{16g}
(\tilde{\partial}_4\hat{h})^2
 - \frac{1}{4g}(\tilde{\partial}_4\hat{h}^a)^2
+ \frac{1}{8g}(\tilde{\partial}_4\hat{h}^{ab})^2
\nonumber \\
   & & \: - 8g(\tilde{\partial}_4^{-1}
\hat{\phi}^{ab})^2 \nonumber \\
   & & \: + \frac{1}{32g} \hat{e}^{ab} \{
-4\tilde{\partial}_4^2 M_1 - (\tilde{\triangle} +
4\tilde{\partial}_4^2) M_2 - 4 \tilde{\partial}_4^2
M_3 \nonumber \\
& & \ \hskip 1cm - (4\tilde{\triangle} + 4
\tilde{\partial}_4^2)
M_4 \}_{(ab)(cd)}
\hat{e}^{cd} \Bigr]\ .
\end{eqnarray}
Note that there does not appear $\varphi$ dependence
in the Lagrangean density in the action
(\ref{action4}).

We now consider the actions and measures for the
ghost fields. By using a vector $k^\mu=(0,0,0,1)$,
the gauge fixing conditions (\ref{gauge2}) and
(\ref{gauge3}) can be rewritten as
follows,
\begin{eqnarray}
    A^a_4 &=& k^\tau A^a_\tau = 0  \label{gauge4} \\
    G^\mu &=& k^\tau \epsilon^{\mu\nu\rho\sigma}
\eta^a_{\nu\rho}
b^a_{\tau\sigma} \approx  k^\tau
\epsilon^{\mu\nu\rho\sigma}
B^a_{\nu\rho} b^a_{\tau\sigma} = 0
\label{gauge5}\ .
\end{eqnarray}
Since the inhomogeneous terms under the $SU(2)$
gauge transformation and the general coordinate
transformation is given by
\begin{eqnarray}
\label{inho1}
\delta A^a_4 &=& \partial_4 \theta^a+\cdots \\
\label{inho2}
\delta b^a_{\mu\nu} &=& \varphi^2 (
\eta^a_{\rho\nu}\partial_\mu\epsilon^\rho
+ \eta^a_{\mu\rho}
\partial_\nu\epsilon^\rho) + \cdots \ ,
\end{eqnarray}
we find that the actions and measures of the
(anti-)ghost fields $( r^a,c^a) $,
$ (p_\alpha , q^\alpha ) $ are given by
\begin{eqnarray}
\label{gact}
   S_{{\rm ghost}} &\sim&  \int d^4x \:
\{ \epsilon^{\mu\nu\rho\sigma}
B^a_{\mu\nu} B^a_{\rho\sigma} k^\tau
r^b \partial_\tau c^b
\nonumber \\
   & &\:\:\: + p_\mu k^\tau
\epsilon^{\mu\nu\rho\sigma}
B^a_{\mu\nu} ( B^a_{\beta\sigma} \partial_
\tau q^\beta +
B^a_{\tau\beta} \partial_\sigma q^\beta) \}
\nonumber \\
    &\sim& \varphi^4\int d^4x   \{
\epsilon^{\mu\nu\rho\sigma}
\eta^a_{\mu\nu} \eta^a_{\rho\sigma} k^\tau
r^b \partial_\tau c^b
\nonumber \\
   & &\:\:\: + p_\mu k^\tau
\epsilon^{\mu\nu\rho\sigma}
\eta^a_{\mu\nu} ( \eta^a_{\beta\sigma}
\partial_\tau q^\beta +
\eta^a_{\tau\beta} \partial_\sigma q^\beta) \}
\end{eqnarray}
%%%%%%%%%%%%%%%%%%%%%%%%%%%%%%%%%%%%%%%%%%%%%%%%%%%%%%%%%%%%%
\begin{eqnarray}
\label{gmeas}
  M_{{\rm ghost}} &\sim &  \int d^4x  \{
\epsilon^{\mu\nu\rho\sigma}
B^a_{\mu\nu} B^a_{\rho\sigma}( \delta r^b
\delta c^b  +
\delta p_\alpha \delta q^\alpha )\}
\nonumber \\
   &\sim& \varphi^4\int d^4x ( \delta r^a
\delta c^a  +
\delta p_\alpha \delta q^\alpha ) \ .
\end{eqnarray}
Here we have kept only quadratic terms with
respect to (anti-)ghost fields.
By using the coordinate system $\{\tilde x^\mu\}$
and redefining the ghost fields
$(p_\alpha, q^\alpha)$ by $ p_\alpha =
\varphi \hat{p}_\alpha $ and
$ q^\alpha= \varphi^{-1} \hat{q}^\alpha $,
the $\varphi$ dependence of measures is
absorbed and we obtain
\begin{eqnarray}
\label{gact2}
   S_{ghost} &\sim& \varphi \int d^4\tilde{x}
\{  \epsilon^{\mu\nu\rho\sigma}\eta^a_{\mu\nu}
\eta^a_{\rho\sigma} k^\tau r^b
\tilde{\partial}_\tau c^b \nonumber \\
   & &\:\:\:  + \hat{p}_\mu k^\tau
\epsilon^{\mu\nu\rho\sigma}
\eta^a_{\mu\nu} ( \eta^a_{\beta\sigma}
\tilde{\partial}_\tau
\hat{q}^\beta + \eta^a_{\tau\beta}
\tilde{\partial}_\sigma
\hat{q}^\beta) \}  \\
   M_{ghost} &\sim&\: \: \: \int d^4\tilde{x}
( \delta r^a
\delta c^a  + \delta \hat{p}_\alpha
\delta\hat{q}^\alpha )\ .
\end{eqnarray}
By redefining $\hat{q}_4 \rightarrow
\hat{q}_4 - \frac{1}{3} \tilde{\partial}_4^{-1}
\tilde{\partial}_a\: \hat{q}^a$,
the second term in the action (\ref{gact2})
is rewritten by
\begin{equation}
\label{gact3}
-2\varphi\int d^4\tilde{x} \{ \hat{p}_4
\:(\:3 \tilde{\partial}_4
\hat{q}_4 )\: + \: \hat{p}_i \:(\:
\tilde{\partial}_4 \hat{q}^i
+  \epsilon^{ia\sigma} \tilde{\partial}_\sigma
\hat{q}^a ) \}\ .
\end{equation}

The one-loop contribution of ghost and anti-ghost
fields
$(p^\alpha, q_\alpha)$
and $(r^a, c^a)$ to the
effective potential can be evaluated by using the
following formula:
\begin{eqnarray}
\label{gdet}
 \int d\beta d\gamma
 \e^{\int d^4\tilde{x}\beta D \gamma }
 &=& \det D =
 \{ \det D \}^{\frac{1}{2}}
 \{ \det D \}^{\frac{1}{2}} \nonumber \\
   &=& \{\det D\}^{\frac{1}{2}}\{
   \det {^TD} \}^{\frac{1}{2}} \nonumber \\
   &=& \e^{\frac{1}{2}\: \Tr\, \ln D\:{^TD}}
   \nonumber \\
&=&  \int d\beta d\bar\beta
 \e^{\int d^4\tilde{x} \beta D\:{^TD} \bar\beta }
\end{eqnarray}
Here $(\beta,\gamma)$ represents $(p^\alpha, q_\alpha)$
and $(r^a, c^a)$. ($D=\varphi\tilde\partial_4$ for
$(p^\alpha, q_\alpha)$ and $D=D^a_b = \varphi
(\tilde{\partial}_4 \delta^a_b
+  \epsilon^{ab\sigma} \tilde{\partial}_\sigma)$ for
$(r^a, c^a)$.)
The explicit form of $D\:{^TD}$ for $(r^a, c^a)$,
which can be
orthogonally decomposed, is given in Appendix B.

\section{Effective Potential}

The one-loop contributions of any fields to
the effective potential take the form of
$ {\rm Tr} \ln \{ A \triangle
 + B \partial_4^2 \} $, which is quartically divergent.
We evaluate this quantity
by using the following regularization:
\begin{eqnarray}
\label{Trln}
 & & F(a,b;\alpha,\beta;\epsilon) \nonumber \\
 &=& \sum_{n_1,n_2,n_3,m=-\infty
\atop(n_1,n_2,n_3,m \ne 0)}^{+\infty}
\ln \{ a( n_1^2 + n_2^2 + n_3^2 ) + bm^2 \}
e^{-\epsilon \{ \alpha (n_1^2 + n_2^2 + n_3^2 )
+ \beta m^2 \} } \ .
\end{eqnarray}
Here $\epsilon$ is a cut-off parameter for the
regularization and the $c$-number coefficients
$\alpha$ and $\beta$
are chosen so as to keep the invariance of the
local symmetry .
By using the following formulae,
\begin{eqnarray}
\label{ln}
\ln Q &=& -\frac{\partial}{\partial \xi}
\frac{1}{\Gamma(\xi)}
\int _0^\infty ds s^{\xi-1} e^{-Qs}
\Big| _{\xi=0}  \\
\label{theta} \vartheta_3 (\upsilon,\tau)
&=& \sum_{n=-\infty}^{\infty} (e^{\tau
\pi i})^{n^2} ( e^{\upsilon\pi i} )^{2n} \ ,
\end{eqnarray}
Eq.(\ref{Trln})  is rewritten by
\begin{eqnarray}
\label{Trln2}
F(a,b;\alpha,\beta;\epsilon)  & =&
-\frac{\partial}{\partial\xi}
\Bigl[ \frac{\epsilon^{\xi}}{\Gamma(\xi)}
\int_0^\infty dt\, t^{\xi-1} \\
& &  \ \times\left\{ \vartheta_3
( 0,\frac{i}{\pi} \epsilon(at+\alpha) )^3
\vartheta_3( 0,
\frac{i}{\pi} \epsilon(bt+\beta)) -1\right\}
\Bigr] \Big|_{\xi=0}\ . \nonumber
\end{eqnarray}
Since the theta function $\vartheta_3
(0,\frac{ix}{\pi})$ has the following
properties,
\begin{equation}
\label{theta2}
\left\{
\matrix{
\lim_{x \to \infty} \vartheta_3
(0,\frac{ix}{\pi}) &=& 1 \cr
\vartheta_3 (0,\frac{ix}{\pi})
&=& (\frac{\pi}{x})^{1/2}
\vartheta_3 (0, \frac{\pi}{x}i) \cr
   &\stackrel{x \to 0}{\rightarrow}
& (\frac{\pi}{x})^
{1/2}\:\:+ {\it o}(x^n) \cr
} \right. \ ,
\end{equation}
we can use the following approximation if we
are interested in the leading term with respect
to the cut-off parameter $\epsilon$:
\begin{equation}
\label{Trln3}
F( a,b;\alpha,\beta;\epsilon) \sim
-\frac{\partial}{\partial
\xi} \left[ \frac{\epsilon^\xi}{\Gamma(\xi)}
\int _0^\infty
dt\: t^{\xi-1}\frac{\pi^2}{\epsilon^2
(at+\alpha)^{3/2}(bt +\beta)^{1/2}}
\right] \Big|_{\xi=0} \ .
\end{equation}
By changing the variable
$ t=\alpha s/a$, we obtain
\begin{eqnarray}
\label{Trln4}
F( a,b;\alpha,\beta;\epsilon) &\sim&
- \frac{\partial}{\partial\xi}
\Bigl[ \frac{\epsilon^{\xi-2}\pi^2}
{\Gamma(\xi) a^{3/2}b^{1/2}} \left(
\frac{\alpha}{a}\right)^{\xi-2} \nonumber \\
& & \ \hskip 0.5cm \times\int_0^\infty
ds\:\frac{s^{\xi-1}}{(s+1)^{3/2}(s+a\beta/b
\alpha)^{1/2} } \Bigr] \Big|_{\xi=0}  .
\end{eqnarray}
The integration
\begin{equation}
\label{int1}
f(x,\xi) \equiv  \int _0^\infty ds
\, s^{\xi-1} \frac{1}{(s+1)^{3/2}(s+x)^{1/2}}
\end{equation}
is given by Gauss' hypergeometric function:
\begin{equation}
\label{int2}
f(x,\xi) = \frac{ \pi (1-\xi)}{\sin(\pi \xi)}
{\cal F}(2-\xi,\frac{1}{2},2;1-x)\ .
\end{equation}
By expanding $f(x,\xi)$ with respect to
$\xi$, we find
\begin{eqnarray}
\label{int3}
   f(x) &=& \left[ \frac{1}{\xi} s^\xi
\frac{1}{ (s+1)^{3/2}
(s+x)^{1/2}} \right]_{s=0}^{\infty} \Big|_{\xi=0}
\nonumber \\
  & &  -\frac{1}{\xi} \int_0^\infty s^{\xi}
\left\{-\frac{3}{2}\frac{1}{(s+1)^{5/2}
(s+x)^{1/2}} - \frac{1}{2}
\frac{1}{(s+1)^{3/2}(s+x)^{3/2}} \right\}
\Big|_{\xi=0} \nonumber \\
 &\approx& -\frac{1}{\xi} \int_0^{\infty} ds
\left(1+\xi \ln s +
O(\xi^n)\right) \{ \:\:\:\:\:\: \}
\Big|_{\xi=0} \nonumber \\
   &\equiv& f_{-1}(x) \xi^{-1} +
f_0(x) \xi^0 + f_1(x)\xi + \cdots
\end{eqnarray}
$f_{-1}$ is given by the incomplete beta function
$B_z(p,q)$. (Explicit form for $f_{-1}$ is given
in Appendix C.
By using Eq.(\ref{int3}), we find the following
expression of $F(a,b;\alpha,\beta;\epsilon)$:
\begin{eqnarray}
      F(a,b;\alpha,\beta;\epsilon) &\sim&
      -\frac{\partial}
{\partial\xi} \left[ \frac{(1+\xi \ln \epsilon
+ {\cal O}(\xi^n))\pi^2  }
{(\frac{1}{\xi} - \gamma + {\cal O}(\xi^n) )
a^{3/2}b^{1/2} }
\left(\frac{\alpha}{a}\right)^{\xi-2}
f(\frac{a \beta}{b \alpha} )
\right] \Big|_{\xi=0}  \nonumber \\
    &=& \frac{-\pi^2}{\epsilon^2
    a^{3/2}b^{1/2} }
\left( \frac{\alpha}{a} \right)^{-2} \nonumber \\
&& \ \hskip 0.5cm \times
\left\{ f_{-1}
(\frac{a \beta}{b \alpha} )
( \ln \frac{\alpha}{a} + \ln \epsilon
+ \gamma )+ f_0
(\frac{a \beta}{b \alpha} ) \right\} \ .
\end{eqnarray}

In order to  determine the coefficients
$\alpha$ and $\beta$, which are the parameters
for the regularization,
we consider actions invariant
under the general coordinate transformation.
In case of the anti-ghost field $r^a$, the invariant action
is given by,
\begin{equation}
\label{ab1}
\int d^4x \sqrt{-g} g^{\mu\nu} \partial_\mu \bar{r}^a
\partial_\nu r^a \sim \int d^4 \tilde{x} \eta ^{\mu\nu}
\tilde{\partial}_\mu \bar{r}^a \tilde{\partial}_\nu r^a \ .
\end{equation}
Here $\bar r^a$ is an anti-self-dual partner of $r^a$
in Eq.(\ref{gdet}).
Since the eigenvalues of
$\tilde\triangle$ and
$\tilde\partial_4^2$ are given by
\begin{eqnarray}
\label{eigen}
   \tilde{\triangle}^2  &=&  -\frac{(2\pi)^2}{L^2\varphi^2}
(n_1^2 + n_2^2 + n_3^2 )  \\
   \tilde{\partial}_4^2 &=& -(2\pi kT)^2 m^2 \ ,
\end{eqnarray}
we find $\alpha$ and $ \beta$ corresponding to $r^a$,
\begin{equation}
   \alpha = \frac{(2\pi)^2}{L^2\varphi^2} \: \: , \: \:
\beta = (2\pi kT)^2 \ .
\end{equation}

In a similar way, we find the actions of
$ (p_\alpha$ , $\bar{p}_\alpha) $, $ B^a_{\alpha\beta}$
and $\phi^{ab}$
%%%%%%%%%%%%%%%%%%%%%%%%%%%%%%%%%%%%%%%%%%%%%%%%%%%%%%%%%%%%%%%%%
\begin{eqnarray}
\label{ab2}
    \int d^4 x \sqrt{-g} g^{\mu\nu}
    g^{\rho\sigma}
D_\mu p_\rho D_\nu \bar{p}_\sigma &\sim&
\int d^4\tilde{x}
\epsilon^{\mu\nu\rho\sigma}
\tilde{\partial}_\mu\hat{p}_\rho
\tilde{\partial}_\nu\hat{\bar{p}}_\sigma \\
    \int d^4x g^{\mu\nu}
    \epsilon^{\alpha\beta\gamma\delta}
\partial_\mu b^a_{\alpha\beta}\partial_\nu
b^a_{\gamma\delta} &\sim&
\int d^4\tilde{x} \eta^{\mu\nu}
\epsilon^{\alpha\beta\gamma\delta}
\tilde{\partial}_\mu \hat{b}^a_{\alpha\beta}
\tilde{\partial}_\nu
\hat{b}^a_{\gamma\delta} \\
    \int d^4x \sqrt{-g} g^{\mu\nu}
    \partial_\mu \phi^{ab}
\partial_\nu\phi^{ab} &\sim& \int d^4
\tilde{x} \eta^{\mu\nu}
\tilde{\partial}_\mu \phi^{ab}
\tilde{\partial}_\nu \phi^{ab} \\
    \int d^4x
    \epsilon^{\alpha\beta\gamma\delta}
\partial_\alpha A^a_\beta
\partial_\gamma A^a_\delta &\sim&
\int d^4\tilde{x}
\epsilon^{\alpha\beta\gamma\delta}
\tilde{\partial}_\alpha \hat{A}^a_\beta
\tilde{\partial}_\gamma \hat{A}^a_\delta
\end{eqnarray}
%%%%%%%%%%%%%%%%%%%%%%%%%%%%%%%%%%%%%%%%%%%%%%%%%%%%%%%%%%%%%%%
and the coefficients $\alpha$ and $\beta$
\begin{equation}
\label{ab3}
 \alpha = \frac{(2\pi)^2}{L^2\varphi^2} \ , \hskip 1cm
 \beta = (2\pi kT)^2 \ .
 \end{equation}
By considering the contributions from all the fields,
we find that the effective potential has the following form:
\begin{equation}
\label{eff}
  V(\varphi,T) \sim \frac{(L\varphi)^3}{\epsilon^2 kT}
\left\{ C_1 \ln (\mu^2 \epsilon) + C_2 \ln \varphi^2 +
c{\rm -number} \right\}\ .
\end{equation}
Here $\mu$ is a renormalization point. Note that
only ghosts can contribute to the coefficients $C_2$.
If $C_2=0$, the quartically divergent part of the
effective potential can be removed by the
renormalization of the cosmological constant
and the potential would not have any physical meaning.
$C_2\neq 0$ means that there is a divergence
which cannot be renormalized and there is a kind of
anomaly. If the cut-off scale has any physical
meaning, {\it e.g.} Planck scale coming from string
theory, the effective potential would give
the following physical implication:
The effective potential has only one non-trivial
($\varphi\neq 0$) minimum. In the low temperature
($T\rightarrow 0$), the minimum is deep and
the metric does not fluctuate.
This means that the metric has a
non-trivial classical value. On the other hand,
in the high temperature, the effective potential is flat
and the fluctuation of the space-time metric is large.
Therefore this effective potential might explain
why there is the universe at present.

\section{Summary}

In the framework of two-form gravity, which
is classically equivalent to Einstein gravity,
we have calculated the one-loop effective potential
for the conformal factor of metric in the finite
volume and in the finite temperature
by choosing a temporal gauge fixing condition.
There appears a quartically divergent term which
cannot be removed by the renormalization of the
cosmological term and we have found a non-trivial
minimum in the effective potential.
If the cut-off scale has any physical meaning,
\eg the Planck scale coming from string theory,
this minimum might explain why the space-time is
generated, \ie why the classical metric has a
non-trivial value.

The two-form gravity theory might be an low energy
effective theory of string theory since the Kalb-Ramond
symmetry, which is characteristic to the two-form
gravity, is
stringy symmetry \cite{kkns}.

\section*{Appendix A. Orthogonal Decomposition for
the Kinetic Terms of $e^{ab}$}

By partially integrating
\begin{equation}
\label{a1}
\{ \epsilon^{bdc} \partial_c \hat{e}^{ad} + \epsilon^{adc}
\partial_c \hat{e}^{bd} + 2\partial_4 \hat{e}^{ab} \}^2 \ ,
\end{equation}
we obtain
\begin{equation}
\label{a2}
    \hat{e}^{ab} \{ -(2\triangle + 4\partial^2_4 )
I_{(ab)(cd)}
+ A_{(ab)(cd)} + B_{(ab)(cd)} \} \hat{e}^{cd} \ .
\end{equation}
Here $I$, $A$ and $B$ are defined by,
\begin{equation}
\left\{
\matrix{
   I_{(ab)(cd)} &\equiv& \frac{1}{2} (\delta^{ac}\delta^{bd} +
\delta^{ad}\delta^{bc} )-\frac{1}{3} \delta^{ab}\delta^{cd} \cr
   A_{(ab)(cd)} &\equiv& \frac{1}{2} (\delta^{ac}\partial_b
\partial_d + \delta^{ad}\partial_b\partial_c + \delta^{bc}
\partial_a
\partial_d+\delta^{bd}\partial_a\partial_c) \cr
    & & \:\:-\frac{2}{3} (\delta^{cd}\partial_a\partial_b +
\delta^{ab}\partial_c\partial_d) +\frac{2}{9}
\delta^{ab}\delta^{cd}
\triangle \cr
   B_{(ab)(cd)} &\equiv& \epsilon^{bce}\epsilon^{adf}\partial_e
\partial_f + \epsilon^{ace}\epsilon^{bdf}\partial_e\partial_f \cr
    & & \:\:+\frac{2}{3} (\delta^{cd} \partial_a\partial_b +
\delta^{ab}\partial_c \partial_d ) -\frac{8}{9}\delta^{ab}
\delta^{cd}\triangle \cr
} \right. \ .
\end{equation}
If we define
\begin{equation}
\label{a3}
C_{(ab)(cd)} \equiv \partial_a\partial_b\partial_c\partial_d
-\frac{1}{3}(\partial_a\partial_b\delta^{cd}\triangle +
\delta^{ab} \partial_c\partial_d\triangle )+ \frac{1}{9}
\delta^{ab}
\delta^{cd}\triangle^2 \ ,
\end{equation}
$I$, $A$, $B$ and $C$ satisfy the following algebra:
\begin{equation}
\label{a4}
\left\{
\matrix{
  A^2=\triangle A + \frac{2}{3}C & &AB=BA=\frac{4}{3}C \cr
  B^2=4\triangle I-4\triangle A + \frac{8}{3}C & & BC=CB
=\frac{2}{3}
\triangle C \cr
  C^2=\frac{2}{3} \triangle^2 C & &CA=AC=\frac{4}{3}
\triangle C \cr
}\right.
\end{equation}
\begin{equation}
\tr\:I=I_{(ab)(cd)} =5\:,\:\:\tr\:A=\frac{10}{3}\triangle\:,\:\:
\tr\:B=-\frac{10}{3}\triangle\:,\:\:\tr\:C=\frac{2}{3}
\triangle^2\ .
\end{equation}
If we define $M_i$'s by
\begin{eqnarray}
\label{a5}
&& M_1 \equiv\frac{3}{2\triangle^2}C\:,\:\:M_2\equiv\frac{1}
{\triangle}(A-\frac{2}{\triangle}C)\:,\:\: M_3\equiv\frac{1}
{\triangle}(B-2A+2\triangle I)\:,\nonumber \\
&& M_4\equiv-\frac{1}{4\triangle}
(B+2A-2\triangle I -\frac{2}{\triangle}C) \ ,
\end{eqnarray}
$M_i$'s satify the equation $M_iM_j=\delta^{ij}M_i$ and
we find
\begin{equation}
\label{a5b}
f(\sum_{i=1}^4 \alpha_i M^i) =
\sum_{i=1}^4 f(\alpha_i) M^i
\end{equation}
for arbitrary function $f(x)$.
By using $M$, we can express $I$, $A$, $B$ and $C$ by
\begin{equation}
\label{a6}
\left\{
\matrix{
  I&=&M_1+M_2+M_3+M_4 \cr
  A&=&\triangle(M_2+\frac{4}{3}M_1) \cr
  B&=&2\triangle(M_3-M_4+\frac{1}{3}M_1) \cr
  C&=&\frac{2}{3}\triangle^2 M_1 \cr
} \right. \ ,
\end{equation}
and we obtain the following orthogonal decomposition
\begin{eqnarray}
\label{a7}
   M&=&-2(\triangle + 4\partial_4^2) I + A + B \nonumber \\
   &=& -4\partial^2_4M_1-(\triangle + 4\partial_4^2)M_2 -
\partial_4^2M_3-(4\triangle +4\partial_4^2)M_4 \\
   & & \:\tr\:M_1=1\:,\:\:\tr\:M_2=2\:,\:\:\tr\:M_3=0\:,\:\:\tr\:
M_4=2 \ .
\end{eqnarray}

%%%%%%%

\section*{Appendix B. Orthogonal Decomposition for
the Kinetic Terms of $r^a$}

In order to evaluate ($ D^a_b \equiv \tilde{\partial}_4
\delta^a_b + \epsilon^{abc}\tilde{\partial_c} $),
\begin{eqnarray}
\label{gdet2}
 \int dp dq\, \e^{\{ p^a D^a_b q^b \}} &=& \det \, D^a_b =
 \{ \det\,D \}^{\frac{1}{2}}\{ \det\,D \}^{\frac{1}{2}} \nonumber \\
   &=& \{\det\ D\}^{\frac{1}{2}}\{
   \det {^TD} \}^{\frac{1}{2}} \nonumber \\
   &=& \e^{\frac{1}{2}\: \Tr\, \ln\, D\:{^TD}}  \ ,
\end{eqnarray}
we orthogonalize $D\:{^TD}$, which is explicitly
given by
\begin{eqnarray}
\label{b1}
    D^a_b\:{^TD^b_c}&=& -\partial_4^2\delta^a_c+\epsilon^{acd}
\tilde{\partial}_4 \tilde{\partial}_d -\epsilon^{acd}\tilde
{\partial}_4 \tilde{\partial}_d + \epsilon^{abd}\epsilon^{bce}
\tilde{\partial}_d\tilde{\partial}_e \nonumber \\
    &=& -\tilde{\partial}_4\delta^a_c + \tilde{\partial}_c
\tilde{\partial}_a -\delta^{ac}\tilde{\triangle} \ .
\end{eqnarray}
If we define
\begin{equation}
\label{b2}
N^{(1)}_{(ab)}=\frac{1}{\tilde{\triangle}}\tilde{\partial}_a
\tilde{\partial}_b\:,\:\:N^{(2)}_{(ab)} = \delta_{ab}-\frac{1}
{\tilde{\triangle}}\tilde{\partial}_a\tilde{\partial}_b \ ,
\end{equation}
$N^{(i)}$'s satisfy the following relation:
\begin{equation}
\label{b3}
N^{(1)}N^{(2)}=0\:,\:\:N^{(1)}N^{(1)}=N^{(1)}\:,\:\:
N^{(2)}N^{(2)}=N^{(2)}
\end{equation}
and we find
\begin{equation}
\label{b4}
D^a_b {^TD^b_c}= -\tilde{\partial}^2_4 N^{(1)}_{(ac)}
-(\tilde{\triangle} + \tilde{\partial}_4^2 )N^{(2)}_{(ac)} \ .
\end{equation}

\section*{Appendix C. The Evaluation
of the Function $f(x,\xi)$}

The incomplete beta function is defined by
\begin{equation}
\label{c1}
    B_z(p,q) \equiv \int_0^z dt\:t^{p-1}(1-t)^{q-1}\ .
\end{equation}
By changing the variable $t=s/(s+1)$, we obtain
\begin{equation}
\label{c2}
B_z(p,q) = \int _0^{\frac{z}{1-z}}ds
= B(p,q) - \int_{\frac{z}{1-z}}^\infty ds\:
\frac{s^{p-1}}{(1+s)^{p+q}} \ .
\end{equation}
The change of the variable
$s\rightarrow z(1+s)/(1-z)$ gives
\begin{eqnarray}
\label{c3}
    B_z(p,q) &=& B(p,q) - \int_0^\infty ds
    \frac{(s+\frac{z}{1-z})^{p-1}}
    {(s+\frac{1}{1-z})^{p+q}} \nonumber \\
    &=& B(p,q) - \int _0^\infty ds (1-z)^q \frac{1}{(s+z)^{1-p}
(s+1)^{p+q}}
\end{eqnarray}
and we find
\begin{equation}
\label{c4}
\int _0^\infty \frac{1}{(s+z)^{1-p}(s+1)^{p+q}}=
\frac{1}{(1-z)^q}\left( B(p,q) - B_z(p,q) \right)
\end{equation}
and
\begin{eqnarray}
\label{c5}
   f_{-1} &=& \frac{1}{(1-x)^2} \Bigl\{ -\frac{3}{2} \left(B(\frac{1}
{2},2) - B_z(\frac{1}{2},2) \right) \nonumber \\
& & \ - \frac{1}{2} \left(B(-\frac{1}
{2},2) - B_z(-\frac{1}{2},2) \right) \Bigr\} \ .
\end{eqnarray}

The function $f(x,\xi)$, which is defined by
\begin{eqnarray}
\label{c6}
   f(x,\xi)&\equiv& \int_0^\infty ds\:
   s^{\xi-1} \frac{1}{(s+1)^{3/2}(s+x)^{1/2}} \nonumber \\
   &=& \int_0^\infty ds\:
   s^{\xi-1} (s+1)^{-2} \left(1+\frac{x-1}{s+1}
 \right) ^{-1/2} \
\end{eqnarray}
can be expressed by Gauss' hypergeometic function.
By using the binomial expansion and the definition
of the beta function
\begin{eqnarray}
\label{c7}
& & (1+x)^\alpha = \sum_{n=0}^\infty \frac{\Gamma
(\alpha+1)}{\Gamma(\alpha-n+1) \Gamma(n+1)}x^n \\
& & B(p,q) = \frac{\Gamma(p)\Gamma(q)}
{\Gamma(p+q)} =\int_0^\infty dt \frac{t^{p-1}}
{(1+t)^{p+q}} \ ,
\end{eqnarray}
we obtain the following expression
\begin{eqnarray}
\label{c8}
    f(x) &=& \sum_{n=0}^\infty \int_0^\infty ds\:s^{\xi-1}\:
(s+1)^{-n-2}
\frac{\Gamma(\frac{1}{2})}{\Gamma(\frac{1}{2})\Gamma(n+1)}(x-1)^n
\nonumber \\
  &=& \sum_{n=0}^\infty
\frac{\Gamma(\xi)\Gamma(-\xi+n+2)}{\Gamma(n+2)}
\frac{\Gamma(\frac{1}{2})}{\Gamma(\frac{1}{2}-n)\Gamma(n+1)}(x-1)^n
\nonumber \\
&=& \frac{\Gamma(\xi)\Gamma(\frac{1}{2})}{\sqrt{\pi}}\sum_{n=0}
^\infty \frac{\Gamma(2-\xi + n)\Gamma(n+\frac{1}{2})}{\Gamma(n+2)}
\frac{(1-x)^n}{n!}\ .
\end{eqnarray}
Here we have used a formula
$\Gamma(z+\frac{1}{2})\Gamma(\frac{1}{2}-z)
=\frac{\pi}{\cos \pi z}$.
By using the definition of Gauss' hypergeometric function
\begin{equation}
\label{c9}
  {\cal F}(\alpha,\beta,\gamma;z) = \frac{\Gamma(\gamma)}
{\Gamma(\alpha)
\Gamma(\beta)}\sum_{n=0}^\infty \frac{ \Gamma(\alpha+n)
 \Gamma(\beta+n)}{\Gamma(\gamma+n)}\frac{z^n}{n}\ ,
\end{equation}
we obtain
\begin{eqnarray}
     f(x)&=&\frac{\Gamma(\xi)\Gamma(\frac{1}{2})}{\pi}
\frac{\Gamma(2-\xi)\Gamma(\frac{1}{2})}{\Gamma(2)} {\cal F}
(2-\xi,\:\frac{1}{2}\:,\:2\:;1-x) \nonumber \\
&=&\frac{\pi(1-\xi)}{\sin(\pi\xi)} {\cal F}
(2-\xi,\:\frac{1}{2}\:,\:2\:;1-x)\ .
\end{eqnarray}
Since ${\cal F}(2,\frac{1}{2},2;1-x)=x^{-1/2}$,
we obtain the following Taylor expansion:
\begin{equation}
\label{c10}
f(x,\xi) =\frac{x^{-1/2}}{\xi} + \left\{ -x^{-1/2} + \partial_\xi
{\cal F}(2-\xi,\:\frac{1}{2},\:2\:;1-x) \Big|_{\xi=0}\right\}
\end{equation}

\end{document}